\documentclass[12pt]{iopart}

\usepackage{graphicx}

\usepackage{epsfig}
\usepackage{subfigure}
\usepackage{psfrag}
\usepackage{amssymb}
\usepackage{iopams}

\usepackage{indentfirst}
\usepackage{color}

\newcommand{\beq} {\begin{equation}}
\newcommand{\eeq} {\end{equation}}
\newcommand{\beqy} {\begin{eqnarray}}
\newcommand{\eeqy} {\end{eqnarray}}

\begin{document}

\title{Harmonic damped oscillators with feedback. A Langevin study}





\author{P {De Gregorio}$^1$, L Rondoni$^{1,2}$, M Bonaldi$^{3,4}$ and L Conti$^5$}

\address{$^1$ Dip. di Matematica, Politecnico di Torino, Corso Duca degli Abruzzi 24, 10129 Torino, Italy}
\address{$^2$ INFN, Sezione di Torino, Via P. Giura 1, 10125, Torino, Italy}
\address{$^3$ Istituto di Fotonica e Nanotecnologie, CNR-Fondazione Bruno Kessler, 38100 Povo, Trento, Italy}
\address{$^4$ INFN, Gruppo Collegato di Trento, Sezione di Padova, 38100 Povo, Trento, Italy}
\address{$^5$ INFN, Sezione di Padova, Via Marzolo 8, I-35131 Padova, Italy}
\ead{paolo.degregorio@polito.it}

\begin{abstract}
We consider a system in direct contact with a thermal reservoir and which, if left unperturbed, is well described by a memory-less equilibrium Langevin equation of the second order in the time coordinate. In such conditions, the strength of the noise fluctuations is set by the damping factor, in accordance with the Fluctuation and Dissipation theorem. We study the system when it is subject to a feedback mechanism, by modifying the Langevin equation accordingly. Memory terms now arise in the time evolution, which we study in a non-equilibrium steady state. Two types of feedback schemes are considered, one focusing on time shifts and one on phase shifts, and for both cases we evaluate the power spectrum of the system's fluctuations. Our analysis finds application in feedback cooled oscillators, such as the Gravitational Wave detector AURIGA.
\end{abstract}


\maketitle

\section{Introduction}

In recent years, several studies (both experimental and theoretical) have analyzed cooling protocols of a signal generated by a thermal source\cite{marquardt,favero,brown,vinante,zhang}. This is achieved using a feedback apparatus attached to the measuring device, and should be intended as an effective cooling from the bath temperature $T$ to the lower effective temperature $T_{eff}$. In essence, a fraction of the signal generated by the thermal source is fed back to counteract the input signal. This in turn has the global effect of decreasing the signal amplitude, as if the measured source has been further cooled. A possible application is that of quantum oscillators\cite{schwab}. Typically, as of today, practical limits prevent one from achieving bath temperatures that are low enough for the quantum nature of the oscillator (e.g. a nanomechanical device) to be clearly detectable. Feedback cooling is used in the attempt to `close the gap'. Another notable application is that of gravitational wave detection\cite{tombesi,ligo}, such as in the AURIGA detector\cite{vinante}, where the feedback is used for technical reasons\cite{vinante2}.

\begin{figure}[t!]
\centering{
\includegraphics[width=0.7\columnwidth]{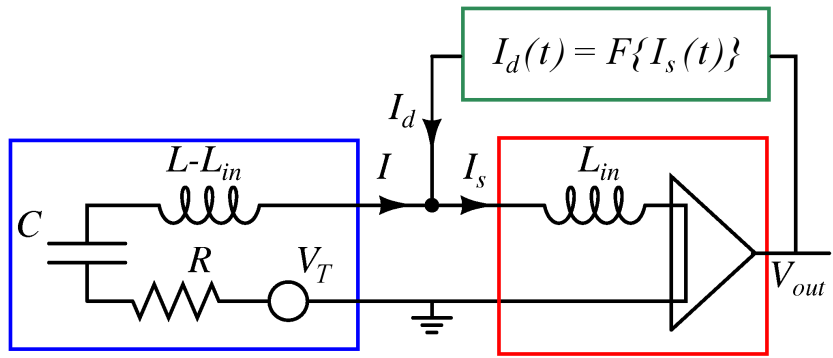}}
\caption{The scheme of a RLC driven by a thermal voltage $V_T$ and connected to a feedback device which `translates' the input current $I_s(t)$ into $I_d(t)$, a functional of $I_s(t)$. A particular realization of this scheme has been already implemented to analyze the output of AURIGA\cite{vinante,bonaldi}.
\label{fig:circuit}}
\end{figure}

Our intent here is to develop a model of the behavior induced by a feedback apparatus on a resonant circuit whose current is driven by a stochastic voltage of thermal origin (see for instance Figure \ref{fig:circuit}).

Some studies have already dealt with similar problems \cite{frank,munakata}, with one crucial difference. The feedback was represented by a delayed potential in a first order Langevin equation, which generates a characteristic oscillatory behavior. We stress that, differently, we focus on the case of a harmonic oscillator. Moreover, the feedback may act on any term of the Langevin equation. We also show the difference between applying the feedback as an ideal delay line in the time domain and a filter in the frequency domain, a distinction that becomes important in many actual realizations.

In the following we shall: i) illustrate how the second order equilibrium Langevin equation should be modified in the presence of feedback terms of all orders in the time derivative; ii) consider the case of an RLC circuit in which the feedback driving is the time derivative of the current coupled to a memory term; iii) derive the power spectra of the current in stationary states when the feedback employs a delayed time-shift (the `digital protocol'), highlighting the fundamental differences that emerge in the case of high, moderate and low quality factors for the electronic oscillator; demonstrating, among others, that the resonance frequency in all cases depends on the feedback gain factor, and that the effect of changing the time shift is markedly different depending on the quality factor; iv) derive the above power spectra if the feedback employs a phase shift (the `analog protocol'), a situation equivalent to that of AURIGA, choosing different cutoff frequencies and discussing similarities and differences with the digital protocol.

\section{From equilibrium to feedback driving}

Let us review the case of a system in equilibrium with a thermal reservoir, which can be described by a one-dimensional equilibrium Langevin equation\cite{kubo},
\beq
\ddot{q} + \gamma \dot{q} + \omega_0^2 q = \eta \label{langequi}
\eeq
where $\eta \equiv \eta(t)$ is a Gaussian noise for which the following FDT relations must hold:
\beqy
&& \langle \eta(t) \rangle = 0 \nonumber \\
&&\langle \eta(t)\eta(t') \rangle =  2 \gamma \langle \dot{q}^2 \rangle \delta(t-t') \label{eta} \\
&&=\frac{2\gamma  k_BT}{m} \delta(t-t') \label{eta_equil}
\eeqy

The quantity $\langle \dot{q}^2 \rangle$ is the square of the observable $\dot{q}(t)$, averaged over the statistical distribution of the noise (therefore a constant in stationary conditions).
Equation (\ref{eta}) is a constraint that holds for any stationary states generated by the dynamics (\ref{langequi}), if $\eta(t)$ is delta-correlated and satisfies the causality relations $\langle \eta(t) q(t') \rangle = 0$ and $\langle \eta(t) \dot{q}(t') \rangle = 0$, for $t > t'$. Equation (\ref{eta_equil}) involves thermodynamic equilibrium properties, presupposing that the quantity $\dot{q}$ represents one quadratic degree of freedom in an underlying Hamiltonian ($m$ being the corresponding `mass' proportionality factor), provided the canonical distribution at temperature $T$ holds.

Equation (\ref{eta_equil}) encompasses a multitude of systems well described by a harmonic oscillator driven by a stochastic force. For the specific case of an RLC circuit coupled with a thermal reservoir, $q(t)$ is the charge and $I(t)= \dot{q}(t)$ is the current. The circuital parameters $R$ (resistance), $L$ (inductance) and $C$ (capacity) are such that: $\gamma= R/L$, $\omega_0^2=1/LC$, $m=L$. The thermal bath thus produces a stochastic voltage $V_T(t)$ with correlations dictated by FDT, i.e. $\langle V_T(t)V_T(t') \rangle = 2 k_BT R \delta(t-t')$. This means that the resistance dissipates all the fluctuations induced by the thermal bath.

We then connect the circuit to a feedback apparatus. The observable quantities at earlier times are continuously stored and fed back into the system at later times. We wait until new stationary states (if they exist) are reached. This corresponds to a new non-equilibrium situation generated by the constant driving influence of the external apparatus. Correspondingly, Equation (\ref{eta_equil}) is modified by the inclusion of the feedback terms. Here we deal with settings that lead to a modification of the dynamics that makes room for memory terms, while still preserving linearity, such as,
\beq
\int_{-\infty}^t  \kappa(t-t') \ddot{q}(t') dt' \, + \, \int_{-\infty}^t \lambda(t-t') \dot{q}(t') dt' + \int_{-\infty}^t \mu(t-t') q(t') dt' = \rho(t) ~ ~ ~ ~   \label{langenonequi1}
\eeq
If $\kappa(t)$ is kernel-invertible, we shall write,
\beq
\ddot{q}(t)  +  \int_{-\infty}^t \gamma(t-t') \dot{q}(t') dt' + \int_{-\infty}^t \beta(t-t') q(t') dt' =N(t)  \label{langenonequi3}
\eeq
We assume that the stochastic processes $\rho(t)=\rho\{\eta(t) \}$ and $N(t)=N\{\eta(t)\}$ are functionals of the same noise-generating source which operates at equilibrium, with no additional source of stochasticity. In other words, we consider a noiseless feedback. $\rho(t)=\rho\{\eta(t) \}$ and $N(t)=N\{\eta(t)\}$ are supposedly known from analyzing the feedback mechanism. Therefore, although Equation (\ref{langenonequi3}) closely resembles well known Langevin equations with memory, it is not guaranteed that FDT-type relations\cite{kubo,vulpiani} are directly applicable here. In typical situations, one arrives at formulas resembling Equation (\ref{langenonequi3}) after a series of qualitative considerations and reasonable physical assumptions about how one should separate the many degrees of  freedom into a deterministic evolution and a stochastic driving\cite{zaccarelli}. At that point, causality conditions are applied, assuming the separation was done correctly. Here we face a rather different non-equilibrium situation, somewhat simpler to some extent. In fact, the functional expression for the cumulative stochastic process $N(t)=N\{\eta(t)\}$ is in principle given once the feedback mechanism is known explicitly. At the same time however, $N(t)$ depends on the past history of $\eta(t)$, rendering the assumption of causality less justified. This issue is somewhat reminiscent of what has been discussed in other works\cite{vulpiani2} and will be clarified later.

\section{Concrete examples - types of feedback}

Let us now proceed with the concrete example of Figure \ref{fig:circuit}. The scheme works by first amplifying the input signal current $I_s(t)$. After an appropriate transformation, a fraction of the original current is then returned to the RLC. For such a transformation we consider two protocols, which we name the `digital' and the `analog' protocols.

The equation describing the dynamics of the circuit is,
\beqy
&&(L-L_{in}) \ddot{q}(t) + R \dot{q}(t) + \frac{1}{C} q(t) = V_T(t) - L_{in}\ddot{q}_s(t)  \nonumber \\
&&I_s(t) = I(t) + I_d(t)  \label{circuit_id}
\eeqy
where $V_T(t)$ is the voltage generated by the thermal bath, and we have neglected the noise generated by the feedback apparatus.

The current $I_d(t)$ is ceded back to the main circuit, i.e. it is the feedback output. It is a functional (here linear) of the current $I_s(t)$, which in turn is both the measured signal and the input of the feedback. The dependence of $I_d(t)$ on $I_s(t)$ is determined by the feedback. It is helpful to express the currents in the frequency domain. Since we consider only stationary states, we can use Fourier transformation. The Fourier transform of $f(t)$ is given by \[ \tilde{f}(\omega) = \int_{-\infty}^{+\infty} e^{i \omega t} f(t) dt. \]
In the {\em digital} protocol $I_d(t)$ is given by $I_s(t)$, scaled by a gain factor $G<1$, and shifted in time by $t_d$.
\beq
I_d(t)=GI_s(t-t_d). \label{digital}
\eeq

The {\em analog} protocol is best described in the frequency domain, and we shall consider a low-pass filter acting on a current proportional to $I_s(t)$,
\beq
\tilde{I}_d(\omega) = \frac{A \Omega}{\Omega -i \omega}  \tilde{I}_s(\omega) \label{analog}
\eeq
with $\Omega$ the angular cut-off frequency and $A$ the gain in the amplification line.

Equations (\ref{digital}) and (\ref{analog}) imply that the current $I_d(t)$ depends on the past history of the circuit, thus revealing that Equation (\ref{circuit_id}) is indeed in the family of Equation (\ref{langenonequi1}).

In either case, in Equation (\ref{circuit_id}) one can eliminate $I_d(t)$, or its Fourier transform $\tilde{I}_d(\omega)$, in favor of $I_s(t)$ or $\tilde{I}_s(\omega)$ respectively.

The digital protocol is a recurrent scheme in theoretical studies. Literally, the feedback is treated as an ideal external `field' which continuously stores and releases the dynamical entries, exploiting an ideal delay line. This model is better suited if one evaluates numerically the temporal evolution with a simple updating rule. In practice, however, it is not obvious whether such ideal systems actually arise in nature, and whether they are easy to be operationally implemented. That the feedback is rather obtained by a filter working in the frequency domain seems a valid alternative. Incidentally, the latter is precisely the way the gravitational wave detector AURIGA operates\cite{vinante}.

The question then arises as to when the two schemes are equivalent to each other, or in what way they differ and under what circumstances. As a matter of fact, in the case in which the currents are almost sinusoidal with angular frequency $\omega_0$ because, e.g., of the oscillator's high quality factor, choosing $t_d=\pi/2 \omega_0$ and $\Omega \ll \omega_0$ leads to $I_d(t) \simeq G \omega_0 q_s(t)$ in both cases (where we took $A\Omega=G\omega_0$). It has already been shown that, to a good approximation, the feedback induces an extra dissipation of the order of $L_{in} G \omega_0$ \cite{vinante,bonaldi}.

Consider now the two protocols.

\section{The digital protocol - $I_d(t)=GI_s(t-t_d)$}

Eliminating $I_s(t)$ from Equation (\ref{circuit_id}), one has,
\beqy
L \ddot{q}(t) - G(L-L_{in}) \ddot{q}(t-t_d) + R \dot{q}(t) -R G \dot{q}(t-t_d) && \nonumber \\
+ \frac{1}{C} q(t) -\frac{G}{C} q(t-t_d)  = V_T(t) - GV_T(t-t_d) && \label{circuit2}
\eeqy

This equation is of the type of Eq. (\ref{langenonequi1}), and shows that the time evolution of $I$ is determined both by the voltage acting directly at time $t$, and by its portion continuously fed back into the system. When $G\equiv 0$, one recovers the equilibrium situation described by Equation (\ref{langequi}), with $\eta(t)=V_T(t)/L$.

We define $x=L_{in}/L$, $y=1-x$, $\gamma=R/L$, $1/LC=\omega_0^2$. As $t$ can be arbitrarily large in a stationary state, we can systematically eliminate $\ddot{q}$ at earlier times in (\ref{circuit2}), to obtain a convergent series (if $|Gy|<1$) like,
\beq
N(t) = \ddot{q}(t) + \gamma \dot{q}(t) + \omega_0^2 q(t)  - \frac{x}{y} \sum_{k=1}^{\infty} (Gy)^k [\gamma \dot{q}(t-kt_d) + \omega_0^2 q(t-kt_d)] \quad \; \label{circuit3}
\eeq
where $N(t)$ is given by
\beq
N(t) =  \big{[} V_T(t) - \frac{x}{y} \sum_{k=1}^\infty (Gy)^k V_T(t-kt_d)\big{]} \frac{1}{L}.  \label{newnoise}
\eeq

Often, the measured quantity is the signal current $I_s(t)$, rather than $I(t)$. Then, eliminating $I(t)$ in favor of $I_s(t)$ one has
\beqy
L \ddot{q}_s(t) - G(L-L_{in}) \ddot{q}_s(t-t_d) + R \dot{q}_s(t) -R G \dot{q}_s(t-t_d) && \nonumber \\
+ \frac{1}{C} q_s(t) -\frac{G}{C} q_s(t-t_d)  = V_T(t) ~ ~ ~ ~ ~ ~ && \label{circuit2bis}
\eeqy
and, analogously to Equation (\ref{circuit3}),
\beqy
N_s(t) &=& \ddot{q}_s(t) + \gamma \dot{q}_s(t) + \omega_0^2 q_s(t) \label{circuit4} \\
&-& \frac{x}{y} \sum_{k=1}^{\infty} (Gy)^k [\gamma \dot{q}_s(t-kt_d) + \omega_0^2 q_s(t-kt_d)] \nonumber
\eeqy
with
\beq
N_s(t) =   \frac{1}{L}\ \sum_{k=0}^\infty (Gy)^k V_T(t-kt_d).  \label{newnoise_s}
\eeq

Equations (\ref{circuit3},\ref{newnoise}) and (\ref{circuit4},\ref{newnoise_s}) are equations of the type of Eq.~(\ref{langenonequi3}). The unknown is the voltage describing the interaction of the circuit with the thermal reservoir. The distribution assigned to $N(t)$ and $N_s(t)$ will induce the probability distributions of $\{I(t),q(t)\}$ and of $\{I_s(t),q_s(t)\}$.

Thus, to make progress we need some assumption about the statistical behavior of the voltage generated by the thermal bath. As customary, we assume that the driving due to the bath is a Gaussian delta-correlated process also when the feedback apparatus is switched on. We thus assume $\langle V_T(t) V_T(t') \rangle  \propto \delta(t-t')$. To fix the constant we assume that - as in equilibrium - all the fluctuations are dissipated by the resistance $R$, so that $\langle V_T(t) V_T(t') \rangle = 2 k_B T R \delta(t-t')$. Alternatively, we shall write $\langle \eta (t) \eta (t') \rangle= 2 v \delta(t-t')$, with $v = k_B T R/L^2=k_B T \gamma /m$. Formally, this implies that $\langle \tilde{\eta}(\omega) \tilde{\eta}(\omega') \rangle = 4 \pi v \delta(\omega + \omega')$. Still by Fourier transformation, one can calculate the power-spectra of the currents $I$ and $I_s$ as,
\beq
S_J(\omega) =  \int_{-\infty}^{+\infty} e^{i \omega t} \langle J(0)J(t) \rangle dt~; \qquad J=I,I_s. \label{spectrum_1}
\eeq

By transformation of both sides of Equation (\ref{circuit3}) and summation over the index $k$, we can express conveniently the formal Fourier transform $\tilde{I}(\omega)$ of the current $I(t)$ (here a stochastic variable) as, \[ \tilde{I}(\omega) \equiv -i \omega \tilde{q}(\omega) = i \omega (1- G y e^{i\omega t_d}) ~ \frac{\tilde{N}(\omega)} {D(\omega)} \] with \[ D(\omega)= \omega^2(1- Gy e^{i \omega t_d}) + i \omega \gamma (1- G e^{i \omega t_d}) - \omega_0^2 (1- G e^{i \omega t_d}). \] From (\ref{circuit4}), an equivalent expression holds for $\tilde{I}_s(\omega)$, with $\tilde{N}_s(\omega)$ in place of $\tilde{N}(\omega)$.

Equation (\ref{newnoise}) yields \[ \tilde{N}(\omega) = \frac{1-G  e^{i \omega t_d}}{1-Gy e^{i \omega t_d}} ~ \tilde{\eta}(\omega) \] while (\ref{newnoise_s}) yields \[ \tilde{N}_s(\omega) = \frac{\tilde{\eta}(\omega)}{1-Gy e^{i \omega t_d}}. \]
The above expressions also indicate that the transfer functions $T(\omega)= \tilde{I}(\omega)/ \tilde{\eta}(\omega)$ and $T_s(\omega)= \tilde{I}_s(\omega)/\tilde{\eta}(\omega)$ are readily available, with \[ T(\omega)= \frac{i\omega(1-G  e^{i \omega t_d})}{D(\omega)} \] and \[ T_s(\omega)= \frac{i\omega }{D(\omega)} \] 

Then, recalling that $J(t) = \int_{-\infty}^{+\infty} e^{-i\omega t} \tilde{J}(\omega) d \omega/2 \pi$, via the hypothesis $\langle \tilde{\eta}(\omega) \tilde{\eta}(\omega') \rangle = 4 \pi v \delta(\omega + \omega')$, (\ref{spectrum_1}) can be written as,

\beqy
S_I(\omega) &=&  \frac{2 v ~ \omega^2 \big{(}1+G^2 - Ge^{i\omega t_d} - G e^{-i \omega t_d} \big{)}}{D(\omega)D(-\omega)} \label{spectrum_3} \\
\nonumber \\
S_{I_s}(\omega) &=& \frac{2 v ~ \omega^2}{D(\omega)D(-\omega)} \label{spectrum_4}
\eeqy
Obviously, the alternative derivation, starting directly from equations (\ref{circuit2}) and (\ref{circuit2bis}), leads to the same result.

We focus on the expression (\ref{spectrum_4}), since $I_s$ is typically the signal that is measured in an experiment\cite{vinante}. The denominator reads explicitly,
\beqy
D(\omega)D(-\omega) = \alpha^2 + G^2 \beta^2 + (1+G^2) \gamma^2 \omega^2 &&  \label{aatotal} \\
+ 2 x G \gamma \omega^3 \sin{(\omega t_d)} - 2 G (\alpha \beta + \gamma^2 \omega^2) \cos{(\omega t_d)} && \nonumber
\eeqy
with
\beq
\alpha = \omega^2-\omega_0^2, \qquad \beta = y ~ \omega^2-\omega_0^2 \nonumber
\eeq

Under the hypothesis that $v = k_B T R/L^2$ as in equilibrium, Equation (\ref{spectrum_4}) with the specification (\ref{aatotal}) constitutes a quantitative estimate for the full spectrum of the current, if the low-noise feedback apparatus has an ideal delay line with gain $G \in (0,1)$.

Regarding the parameters of the circuit, we analyze different possibilities. First, we consider a high quality factor $\gamma \ll \omega_0$, i.e. $R \sqrt{C/L} \ll 1$. We further assume that $\gamma \ll xG \omega_0 \ll \omega_0$. If the feedback operates as an effective cooling, or damping, it is appropriate to tune the delay time $t_d$ by setting $t_d \omega_0 \simeq \pi /2$. $\omega_0$ is the natural resonance frequency of the oscillator, defined as $\omega_0=1/\sqrt{LC}$.

In the proximity of the resonance frequency, $t_d \omega \simeq \pi /2$ implies that $\cos{(\omega t_d)} \simeq 0$ and $\sin{(\omega t_d)} \simeq 1$. The term cubic in $\omega$ in Equation (\ref{aatotal}) is also negligibile. After neglecting the smallest terms, one verifies that the denominator in (\ref{spectrum_4}) is approximately,
\beqy
D(\omega)D(-\omega) \simeq \big{(}1+y^2G^2 \big{)} \omega^4 - 2 \big{(} 1+yG^2 \big{)} \omega_0^2 \omega^2 + \big{(} 1+G^2 \big{)} \omega_0^4
\eeqy
which implies a power-spectrum of a Lorentzian form,
\beq
S_{I_s}(\omega) = \frac{2 Z \omega^2}{(\omega^2-\omega_r^2)^2 + \mu^2 \omega^2} \label{lorentz}
\eeq
The resonance frequency of the circuit, i.e. the frequency that maximizes this power spectrum, is $\omega_r$. At equilibrium, $Z=v=k_B T R/L^2$, $\mu=\gamma= R/L$ and $ \omega_r = \omega_0 =1/\sqrt{LC}$. With the feedback active, both the resonance frequency and the effective dissipation become dependent on the feedback gain. More specifically, $\gamma \ll xG \omega_0 \ll \omega_0$ entails
\beqy
\omega_r^4 &=& \frac{1+G^2}{1+y^2G^2} \, \omega_0^4 \label{omega4} \\
\nonumber \\
\mu^2 &=& 2 \, \bigg{(}  \sqrt{\frac{1+G^2}{1+y^2G^2}} - \frac{1+yG^2}{1+y^2G^2} \bigg{)} \, \omega_0^2 \nonumber \\
\nonumber \\
Z&=& \frac{v}{1+y^2G^2} \nonumber
\eeqy
The expressions for the resonance frequency and for the damping factor can be further simplified when $1-y=x=L_{in}/L \ll 1$ and $G \ll 1$, leading to,
\beq
\omega_r^2 \simeq \Big{(} 1+\frac{xG^2}{1+G^2} \Big{)} \, \omega_0^2, \qquad \mu \simeq xG\omega_0 \label{omega_mu_appr}
\eeq

Thus, the resonant frequency explicitly varies with the feedback gain, albeit very slightly, a feature that may have contacts with other situations\cite{metzger}. The power spectrum of the measured current is essentially equivalent to that of a similar circuit at equilibrium, with resonance in the proximity of $\omega_0$ but with damping increased by a relative factor approximately $xG\omega_0/\gamma=L_{in}G\omega_0/R$ and temperature lowered by the same factor. The fact that the feedback can indeed act as an effective additional damping is well established\cite{marquardt,favero,brown,vinante,zhang}.

\begin{figure}[t!]
\centering{
\includegraphics[width=0.6\columnwidth,height=!,angle=-90]{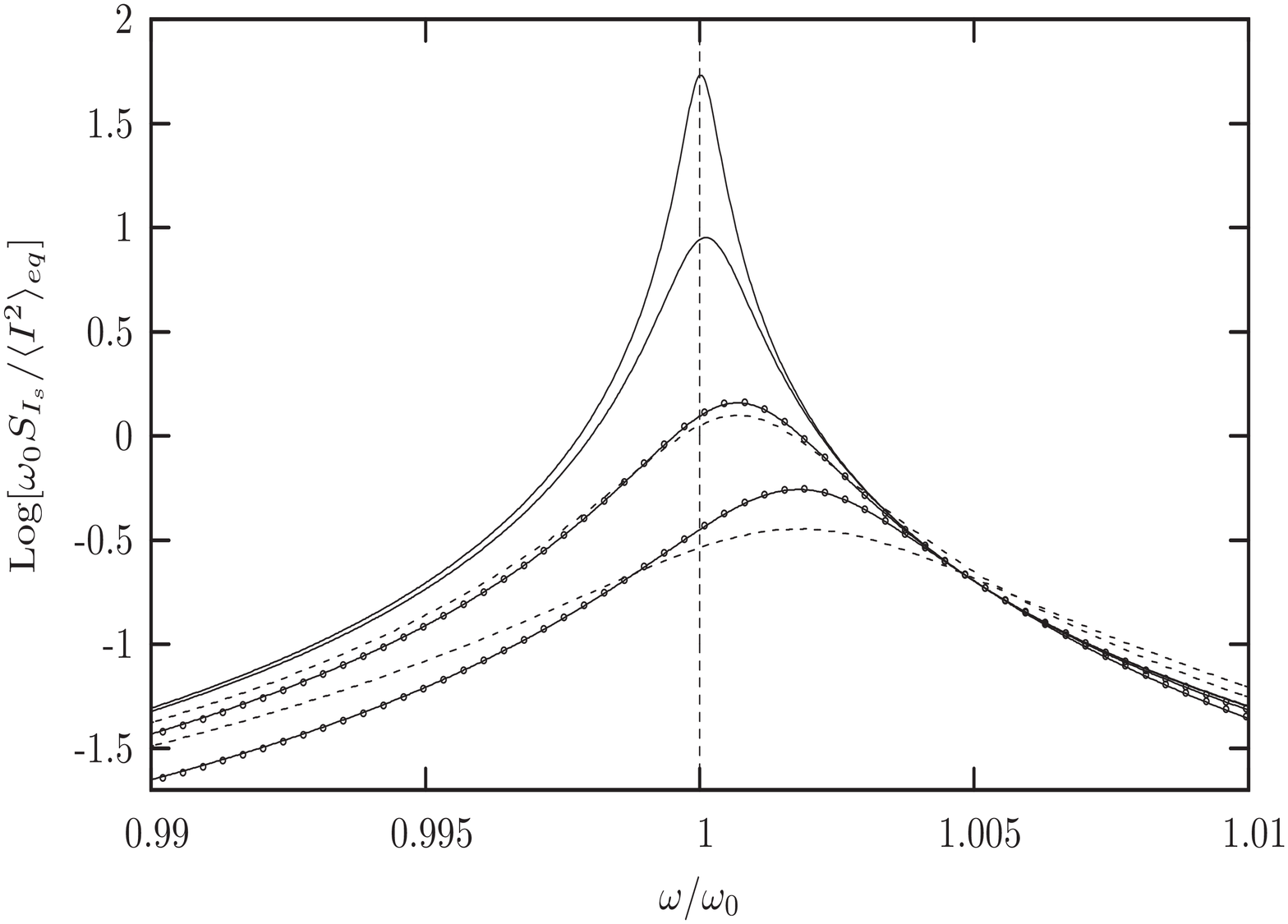}}
\caption{Effective cooling at very high quality factors. Vertical axis in logarithmic scale in base $10$. The power spectrum of the signal current $I_s$ as a function of the normalized frequency $\omega/\omega_0$. $\langle I^2 \rangle_{eq}$ is the mean squared current at equilibrium. Data are for $\omega_0/\gamma = Q=10^{5}$; $x=L_{in}/L=10^{-2}$; $t_d \omega_0=\pi/2$. From top to bottom; the curves are for $G=0.06$; $G=0.15$; $G=0.4$; $G=0.75$. The solid lines represent Equation (\ref{spectrum_4}) for the four settings. The ($\circ$) points are from (\ref{lorentz}) and (\ref{omega4}), showing almost perfect overlap. The dashed lines represent the further approximations of Eqs. (\ref{omega_mu_appr}). For the two top curves at $G=0.06$ and $G=0.15$ we only show one line because the three expressions are indistinguishable. Vertical line is set at $\omega=\omega_0$ as a guide for the eye for the slight shift of the resonance $\omega_r$ away from $\omega_0$ with varying $G$. 
\label{fig:ps1}}
\end{figure}

In Figure \ref{fig:ps1} we show the power spectra (\ref{spectrum_4}), using $Q=\omega_0 / \gamma \equiv \sqrt{L}/(R\sqrt{C}) = 10^5$ and $x \equiv L_{in}/L=10^{-2}$. We choose four different values of $G$: $0.06$, $0.15$, $0.4$ and $0.75$. We compare the full expression (\ref{spectrum_4}) with the curves deduced from the Lorentzian (\ref{lorentz}), using respectively the set of values of Equation (\ref{omega4}) (the $\circ$ points) and the set of Equation (\ref{omega_mu_appr}) (the dashed lines). For $G=0.06$ and $G=0.15$ the three curves are almost identical, but their differences can be appreciated for $G=0.4$ and $G=0.75$. We conclude that (\ref{spectrum_4}) approximates very well a Lorentzian with parameters (\ref{omega4}), and to a lesser degree is also consistent with (\ref{omega_mu_appr}).

\begin{figure}[t!]
\centering{
\includegraphics[width=0.6\columnwidth,height=!,angle=-90]{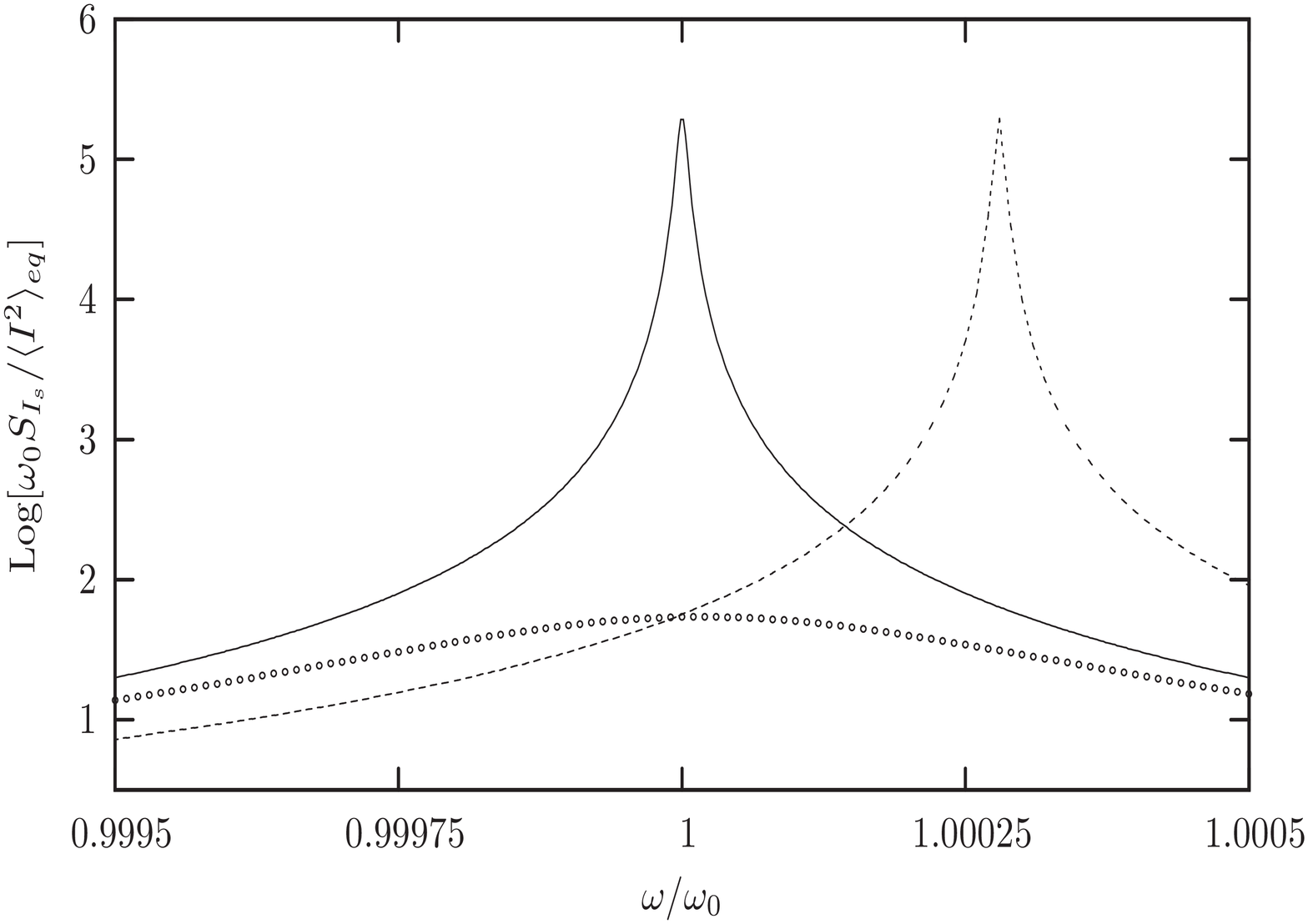}}
\caption{The power spectrum of the current $I_s$ at $Q=10^{5}$, for two different choices of the delay time, against the equilibrium spectrum. Coordinate axes are as in Figure \ref{fig:ps1}. Represented are the equilibrium curve (the solid line), and two curves with active feedback with $x=10^{-2}$; $G=0.06$. The $\circ$ points show the shift $t_d \omega_0=\pi/2$. The dashed line is for $t_d \omega_0=\pi$. 
\label{fig:ps1b}}
\end{figure}

In Figure \ref{fig:ps1b}, we show the peculiar behavior of the spectrum derived from Equations (\ref{spectrum_4}) and (\ref{aatotal}), when we set the delay time $ t_d \omega_0 = \pi$. We compare the two results with the spectrum obtained in the absence of feedback and with that obtained with $ t_d \omega_0 = \pi/2$. Rather than the over-damping characteristic of $t_d \omega_0 = \pi /2$, $t_d \omega_0 = \pi$ entails a marked shift of the resonance frequency with respect to equilibrium.

\begin{figure}[t!]
\centering{
\includegraphics[width=0.6\columnwidth,height=!,angle=-90]{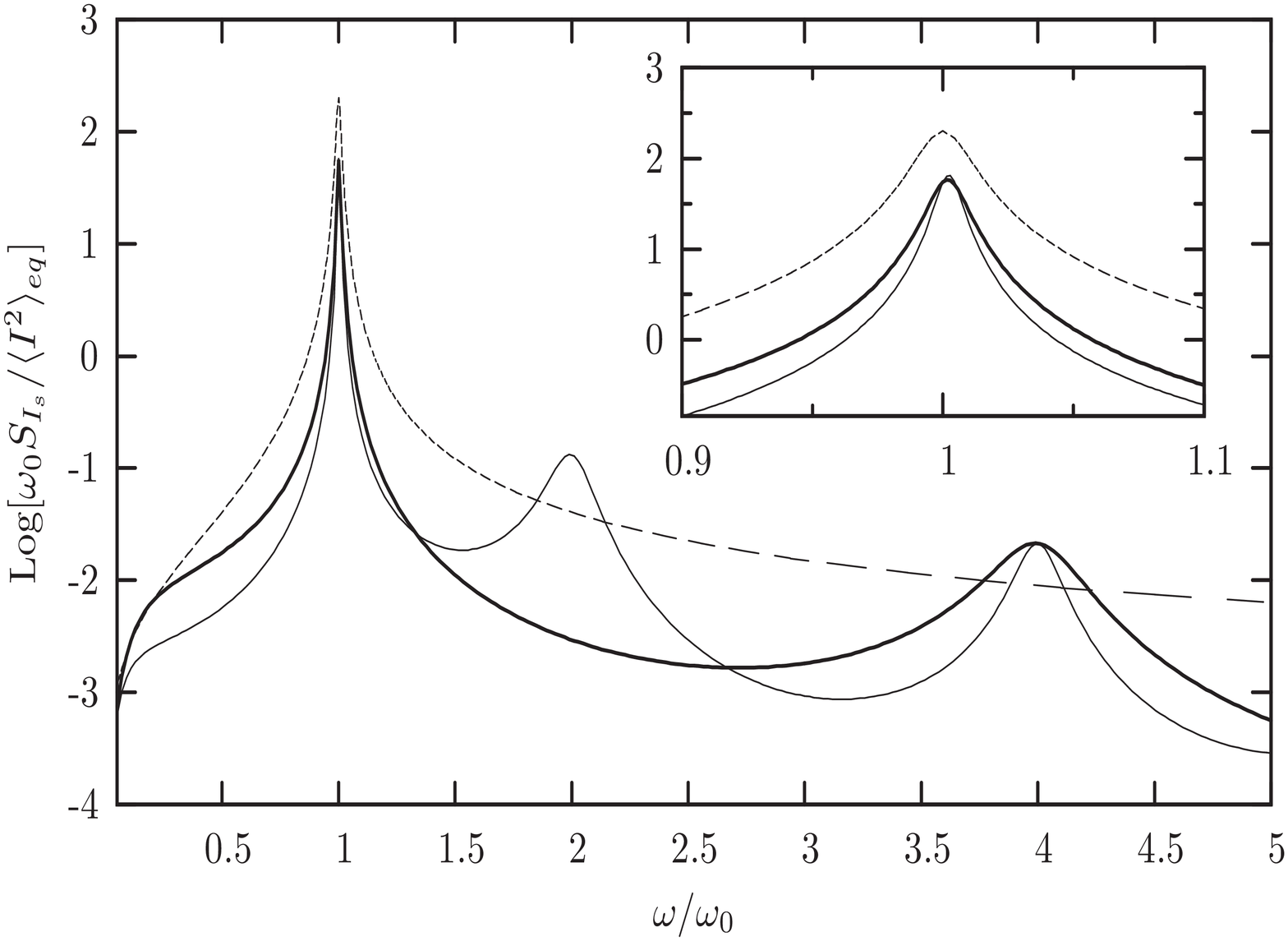}}
\caption{Effective cooling at moderate quality factors. Development of secondary peaks. Vertical axis in logarithm scale in base $10$. The power spectrum of the signal current $I_s$ as a function of the normalized frequency $\omega/\omega_0$. Data are for $\omega_0/\gamma = Q=10^{2}$; $x=L_{in}/L=10^{-2}$; $G=0.75$. Thick solid line shows $t_d \omega_0=\pi/2$; thin solid line is for $t_d \omega_0=\pi$; dashed line is the equilibrium case (absence of feedback). Inset: Zoom in around $\omega_0$.
\label{fig:ps2}}
\end{figure}
\begin{figure}[h!]
\centering{
\includegraphics[width=0.6\columnwidth,height=!,angle=-90]{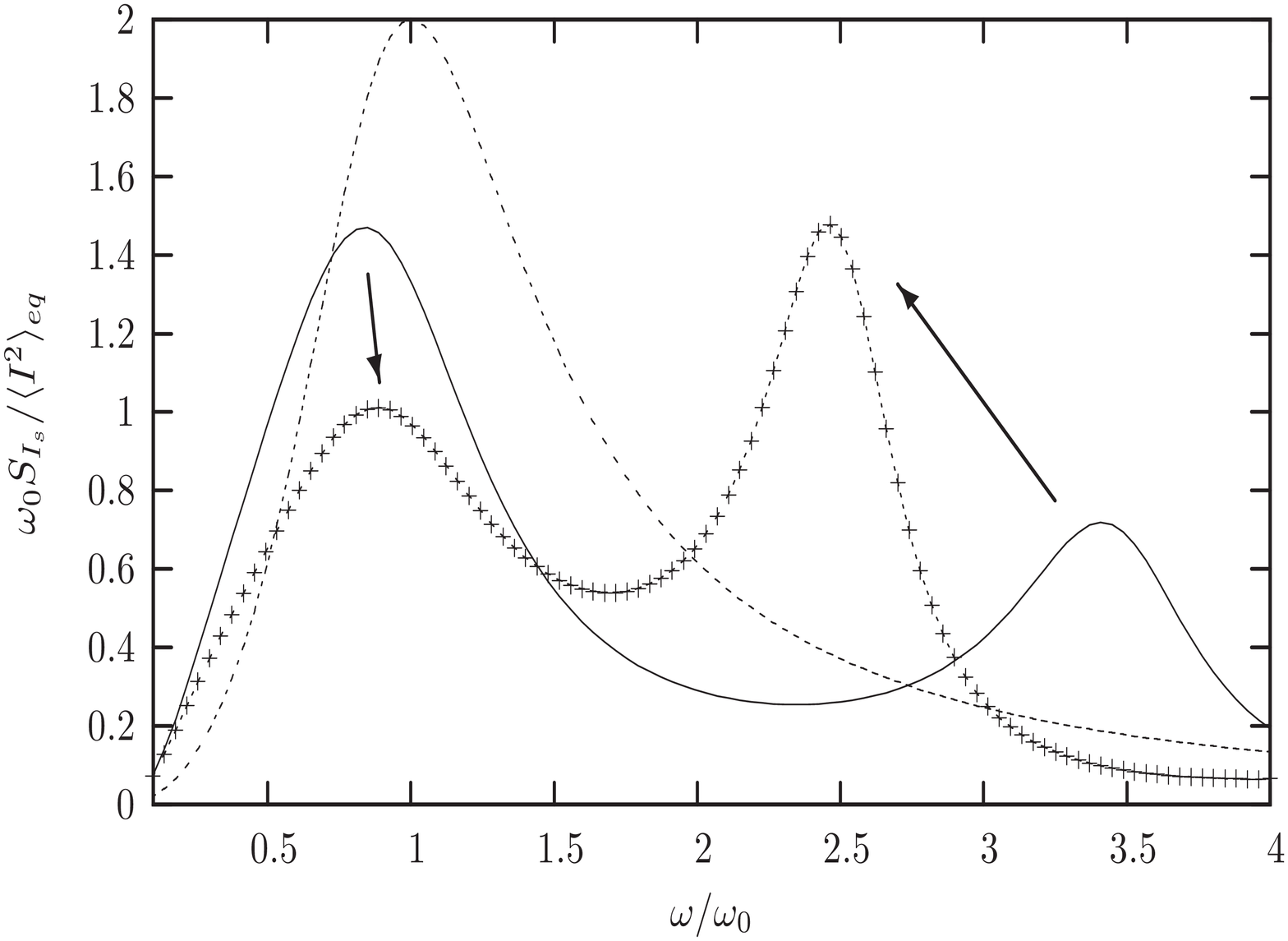}}
\caption{Effective cooling at low quality factors; the `bimodal' trend. The power spectrum as a function of the normalized frequency $\omega/\omega_0$. Data are for $\gamma/\omega_0 = Q=1$; $x=L_{in}/L=10^{-2}$; $G=0.75$. Solid line shows $t_d \omega_0=0.58 \pi$. $+$-dashed curve is for $t_d \omega_0 =0.80 \pi$. The dashed line is the curve at equilibrium. The arrows indicate the trend of the two peaks with increasing $t_d$. The two peaks do not overlap during the `transition'.
\label{fig:ps3}}
\end{figure}

The whole picture changes if we start decreasing the value of $Q$. For example, let $Q=100$. As shown in Figure \ref{fig:ps2}, in this case two features develop. A set of secondary peaks appears, showing that the spectrum ceases to be purely a Lorentzian and becomes an oscillatory function of its argument $\omega$. The reduction of the spectrum amplitude around the natural resonance frequency $\omega_0$ is now accompanied by a slight amplification in correspondence of other frequencies. Furthermore, if we restrict our attention to the close proximity of the equilibrium resonance frequency (the inset of Figure \ref{fig:ps2}), the setting $t_d\omega_0=\pi$ compares very differently than before to either the equilibrium case or the setting $t_d \omega_0=\pi/2$. See for example Figure \ref{fig:ps1b} and our previous discussion. These two features show that in these conditions, despite $Q$ still being relatively high, the feedback drives the system away from its equilibrium configuration in a characteristic way, quite different from the previous case.

Finally, $Q=1$ sets a radically different scenario. Now the dissipation is so high that the feedback has a more dramatic impact on the frequency profile. As we see in Figure \ref{fig:ps3}, changing the delay $t_d$ makes the resonance frequency $\omega_r$ (that at which the spectrum has a maximum) to jump discontinuously from one value to another, for the same gain factor $G$. This is due to the fact that the closest secondary peak increases in amplitude with growing $t_d$.

One can explore a variety of circuital parameters, assuming the general expression for the power-spectrum (\ref{spectrum_4}) holds. For very high $Q$, we have identified the Lorentzian (\ref{lorentz}) and the parameters $Z$, $\omega_r$ and $\mu$. This is a {\it quasi} equilibrium scenario, although radically different from it in a thermodynamic perspective (see \cite{bonaldi}). As the quality factor decreases, we have seen some special features develop. The feedback that operates at a fixed frequency, as in the digital protocol, radically changes the frequency profile of the oscillator. The question is, how much of this richness is an artifact of the feedback operating with a fixed delay time?

In the next section we shall try to answer to this question, by considering a different feedback mechanism.

\section{The analog protocol - $\tilde{I}_d(\omega) = \frac{A \Omega}{\Omega -i \omega}  \tilde{I}_s(\omega)$}

As we have said, an alternative way to implement the feedback protocol is to let the amplified current $AI_s(t)$ be passed through a low-pass normalized filter of cutoff frequency $\Omega \ll \omega_0$. In this case, Equation (\ref{circuit_id}) is best managed directly in the frequency domain. We eliminate $\tilde{q}(\omega)$ and $\tilde{I}(\omega)$ in favor of $\tilde{q}_s(\omega)$ and $\tilde{I}_s(\omega)$, and in order to do that we introduce \[ \tilde{I}(\omega) = \tilde{I}_s(\omega) - \tilde{I}_d(\omega) = \Big{(} 1- \frac{A \Omega}{\Omega-i\omega} \Big{)} \tilde{I}_s(\omega). \]
By Fourier transforming Equation (\ref{circuit_id}) and using the usual notation, one finds,
\beqy
&&B(\omega) ~ \tilde{q}_s(\omega) = (i \omega - \Omega) ~ \tilde{\eta}(\omega) \label{qs_frequency} \\ && \nonumber \\
&&B(\omega) = \omega^2[(1-yA)\Omega-i\omega] \nonumber \\
&& ~ ~ + ~ i \gamma \omega [(1-A)\Omega-i\omega] - \omega_0^2 [(1-A)\Omega-i\omega] \nonumber
\eeqy
Since $\tilde{I}_s(\omega) = -i\omega \tilde{q}_s(\omega)$, Equation (\ref{qs_frequency}) leads once again to an expression of the transfer function for the current, $T_s(\omega) = \tilde{I}_s(\omega) / \tilde{\eta}(\omega)=(\omega^2+i\omega\Omega)/B(\omega)$, and the power spectrum now reads,
\beq
S_{I_s}(\omega) = \frac{2 v ~ \omega^2 ~ (\omega^2+\Omega^2)}{B(\omega) ~ B(-\omega)} \label{spectrum_freq}
\eeq
with
\beq
B(\omega)B(-\omega) = \omega^2(\omega^2-F^2)^2 +(a\omega^2-b^3)^2
\eeq
and
\beqy
F^2 &=& \omega_0^2 + \gamma \Omega (1-A) \nonumber \\
a &=& (1-yA)\Omega + \gamma \nonumber \\
b^3 &=& \Omega \omega_0^2(1-A) \nonumber
\eeqy
Unlike the expression (\ref{spectrum_4}), the power spectrum (\ref{spectrum_freq}) is represented by the ratio of two polynomials in $\omega^2$, a quadratic one and a cubic one. $\Omega^2$ in the numerator may be neglected compared to $\omega^2$ around the resonance, if $\Omega \ll \omega_0$. Simplifications in the denominator will depend on the set of parameters $A$, $y=1-x$ and $\gamma$ that are chosen, and to which $\Omega/\omega_0$ must be compared.

\begin{figure}[b!]
\centering{
\subfigure[$\Omega/\omega_0=0.2$]
{\label{fig:ps4a}
\includegraphics[width=0.52\columnwidth,height=!,angle=-90]{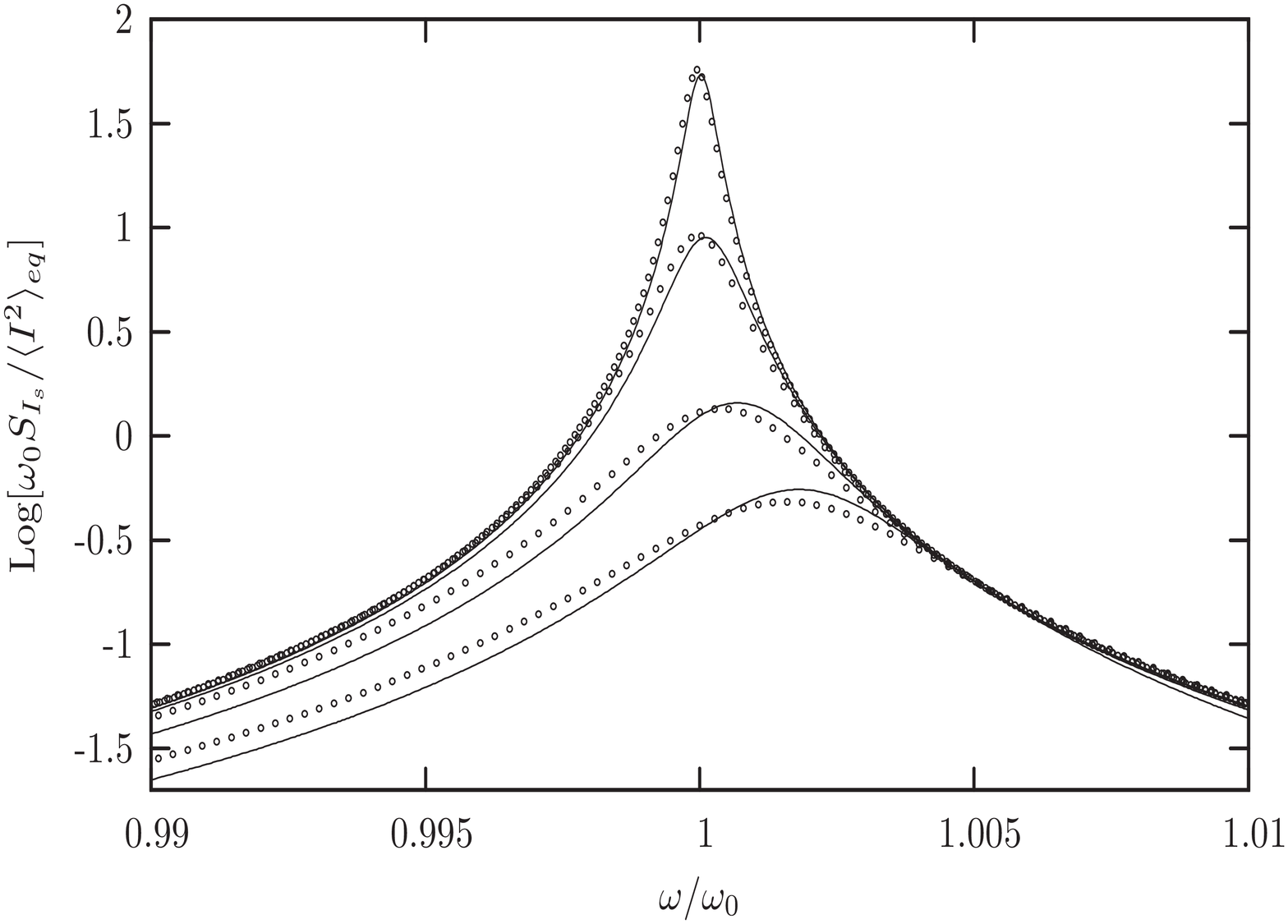}
}
\subfigure[$\Omega/\omega_0=10^{-3}$]
{\label{fig:ps4b}
\includegraphics[width=0.52\columnwidth,height=!,angle=-90]{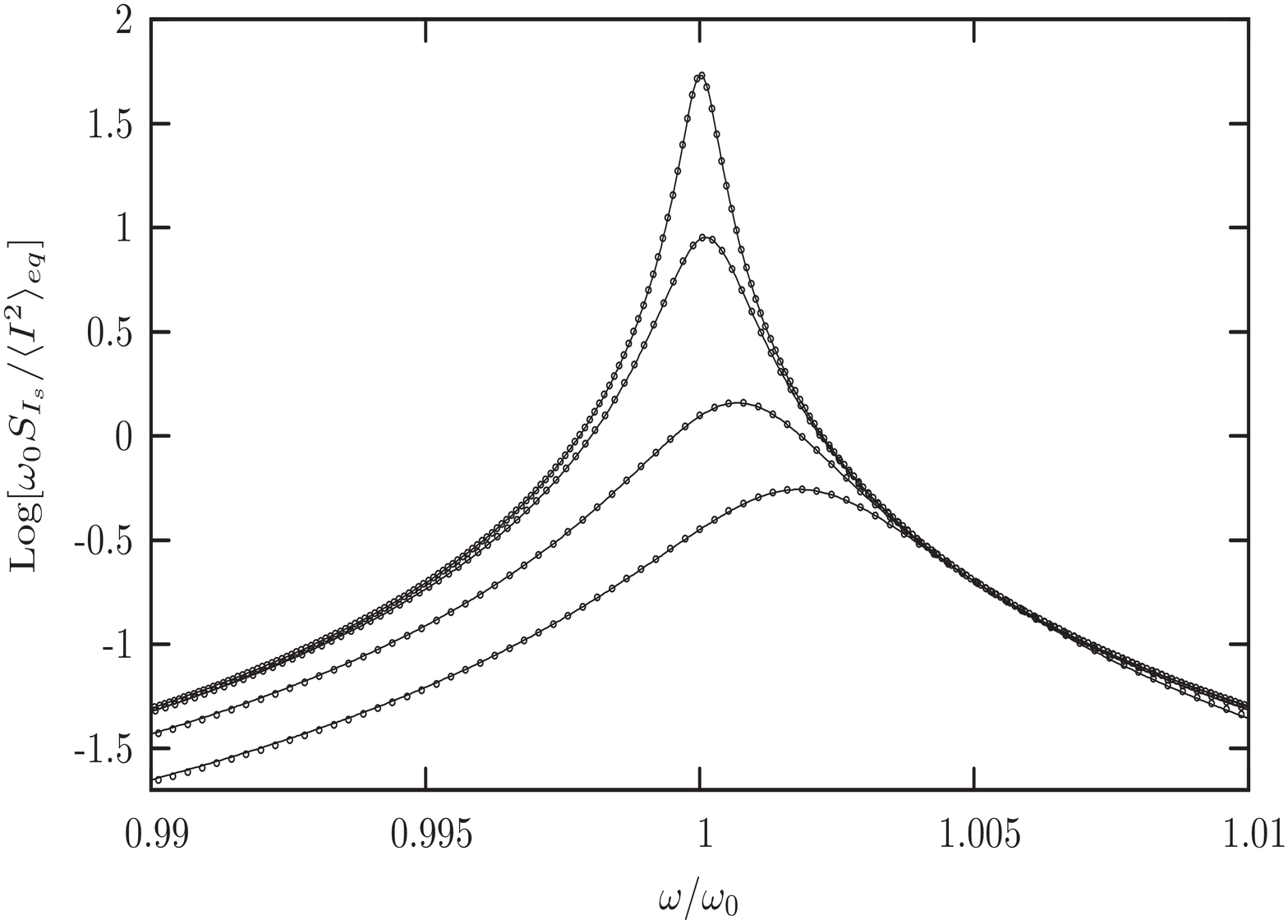}
}}
\caption{Comparison of analog protocol Equation (\ref{spectrum_freq}) (the $\circ$ points) and digital protocol Equation (\ref{spectrum_4}) with $t_d\omega_0=\pi/2$ (solid lines). The normalized power spectrum in Logarithmic scale in base $10$ for two different settings of $\Omega/\omega_0$: (a) $=0.2$: (b) $=10^{-3}$. $Q=10^5$; $x=L_{in}/L=10^{-2}$. For each plot, settings from top to bottom: $G=0.06$, $G=0.15$, $G=0.4$, $G=0.75$. $A$ in Equation (\ref{spectrum_freq}) has been varied in all curves to implement the constraint $A\Omega \equiv G \omega_0$.\label{fig:ps4}}
\end{figure}

To link the present formalism to the previous one, it is appropriate to take $A \Omega \equiv G \omega_0$, because in the limit of high $Q$, small $\Omega/\omega_0$ and for $t_d \omega_0=\pi/2 $, the definition $\tilde{I}_d(\omega) = \frac{A \Omega}{\Omega -i \omega}  \tilde{I}_s(\omega)$ implies $I_d(t) \simeq \frac{A \Omega}{\omega_0} I_s(t-t_d)$. For the digital protocol we had $I_d(t) = GI_s(t-t_d)$, and thus the identification.

In Figure \ref{fig:ps4} we compare Equation (\ref{spectrum_4}) and Equation (\ref{spectrum_freq}) in the two cases of $\Omega/\omega_0$ large and small, as we are now in the position to compare the two protocols.

In Figure \ref{fig:ps4a}, Equation (\ref{spectrum_freq}) with $\Omega/\omega_0=0.2$ is compared with Equation (\ref{spectrum_4}), with $A=0.3$, $A=0.75$, $A=2$ and $A=3.75$ ($G=0.06$, $G=0.15$, $G=0.4$ and $G=0.75$ respectively). Closer inspection reveals that with the analog protocol the shift of the resonance frequency, quantified by $\Delta \omega \equiv \omega_r-\omega_0$, is equal to $0$ at the two extremes of the interval $A \in [0,1]$, and is both non monotonic and negative inside the interval. A minimum of approximately $\Delta \omega/\omega_0=-5 \times 10^{-5}$ is attained in the vicinity of $A=0.5$, before that quantity becomes monotonically increasing with $A$. In Figure \ref{fig:ps4b} we used $\Omega/\omega_0=10^{-3}$ for the analog protocol, again in comparison with Equation (\ref{spectrum_4}), for the values $A=60$, $A=150$, $A=400$ and $A=750$ (i.e. $G=0.06$, $G=0.15$, $G=0.4$ and $G=0.75$, as before). In this case, the smallness of $\Omega/\omega_0$ leads to an almost perfect overlap for all values of $G$. For this choice of $\Omega/\omega_0$, the shift $\Delta \omega$ appears to be consistently monotonic with growing $A$ (and therefore $G$), in analogy with the digital protocol.

As it transpires, the two systems are very similar for extremely high quality factors (here $Q=10^{5}$), provided $\Omega$ is sufficiently small. In fact, one could show that they overlap completely when we take $\Omega/\omega_0 \rightarrow 0$ (while keeping the ratio $A\Omega/\omega_0$ constant) and we tune the gain factor by adjusting the amplification $A$ to achieve the required gain $G$. This is somewhat surprising if we compare the expressions (\ref{spectrum_4}) and (\ref{spectrum_freq}), and it demonstrates the fact that an ideal filter has exactly the same effect on the current as that of a feedback with an ideal delay line.

\begin{figure}[h!]
\centering{
\includegraphics[width=0.6\columnwidth,height=!,angle=-90]{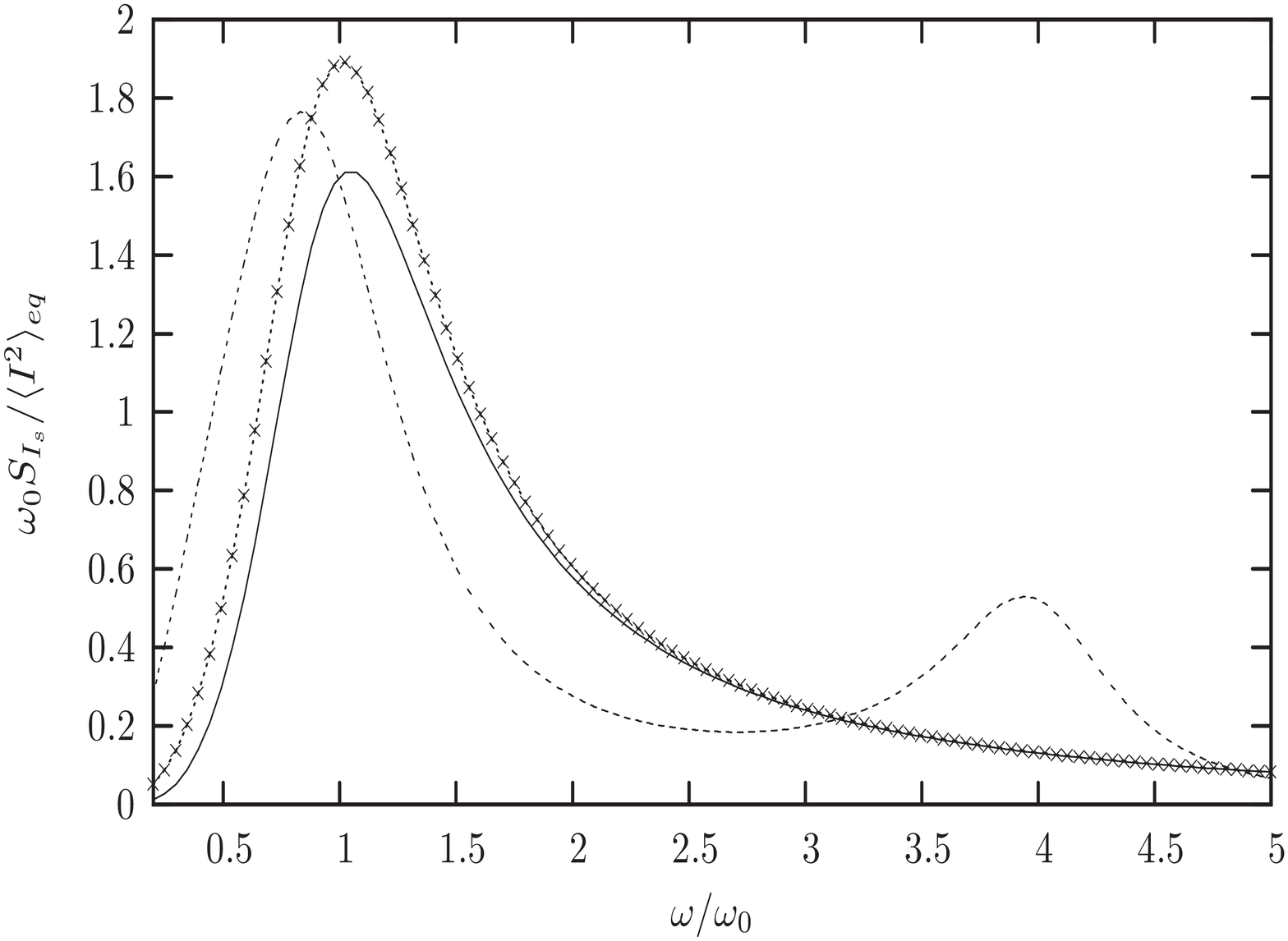}}
\caption{Comparison of Equation (\ref{spectrum_freq}) and Equation (\ref{spectrum_4}) at low quality factor. The normalized power spectrum is shown in linear scale. $Q=1$; $x=L_{in}/L=10^{-2}$; $G=0.5$. Solid line: phase-shifted feedback protocol Equation (\ref{spectrum_freq}), with $\Omega/\omega_0=10^{-3}$. X-dotted curve: Equation (\ref{spectrum_freq}) with $\Omega/\omega_0=0.2$. Dashed line: time-delayed feedback protocol Equation (\ref{spectrum_4}), with $t_d \omega_0=\pi/2$. $A$ in Equation (\ref{spectrum_freq}) has been varied in all curves to implement the constraint $A\Omega \equiv G \omega_0$.
\label{fig:ps5}}
\end{figure}

Finally, as we show in Figure \ref{fig:ps5}, the qualitative difference between the two feedback protocols starts to emerge when we consider low quality factors $Q$ (for example equal to $1$). From the power spectra derived from Equation (\ref{spectrum_freq}) one does not see the same features that can be seen for the digital protocol (Figure \ref{fig:ps3}), with the development of secondary peaks. With the analog protocol, no new characteristic oscillatory behavior emerges. The power spectra here are much more similar in shape to one another. 

Another feature that one is able to appreciate from Figure \ref{fig:ps5} is that for a given $G$ a higher cooling efficiency is obtained by choosing $A$ large and $\Omega/\omega_0$ small.

\section{A case study - the AURIGA detector}

The GW detector AURIGA is an example of the high $Q$ case with the analog protocol. The detector can be modeled by three coupled low-loss resonators\cite{tricarico}: two mechanical ones (the bar and a plate of the capacitive transducer) and an LC electrical one\cite{baggio}. Their dynamics is described by three normal modes at separate frequencies, and each mode is modeled as a RLC series electrical oscillator. Each mode is driven by Johnson noise and the resulting current is fed back to reduce its jitter below that set by the thermal bath\cite{vinante}. The lowest frequency mode out of the 3 is well separated in frequency from the other two and is the best approximation of a single oscillator\cite{vinante,bonaldi}. Here $\Omega/\omega_0=0.23$, $Q \simeq10^6$, $x \simeq 10^{-2}$ and $G \simeq 1.7 \times 10^{-2}$. Two different approaches have anticipated this scenario. One shows that, as in equilibrium, the power spectrum of the current is Lorentzian like, but with effective damping much higher (and effective temperature much lower) than in equilibrium\cite{vinante}. The other approach verified that the probability distributions of the thermodynamic observables can be very well fitted by an equilibrium-like Langevin equation, with a modified temperature and modified damping governing the thermal driving\cite{bonaldi}. Given the high $Q$, the work in \cite{bonaldi} approximated the analog protocol used in the experiment with a digital protocol in analyzing  the stochastic time evolution of the system. Here we show that the approximation is valid also in the frequency domain as shown in Figure \ref{fig:ps4a} and related discussion. At the same time, the present approach highlights some nontrivial features pertaining to the effect of the feedback on the resonance frequency.



\section{Conclusions}

While it is customary to approach the theoretical analysis by working in the time domain with ideal delay lines, we have shown that in some situations there is a distinction between obtaining an active feedback via this method and via filtering the frequencies. For high quality factors the difference is subtle, and can be nullified by appropriate choices of the settings. Conversely, there are irreducible differences for the responses to the feedback of harmonic oscillators with low quality factors.

Our analysis improves our understanding of some nonequilibrium aspects of oscillators, currently of much interest in large deviations theory\cite{fluctuations}. Lack of a complete thermodynamic description of the source and the feedback apparatus is still a problem in these contexts, since they are mostly concerned about dissipation, work and exchanged heat\cite{gonnella_harris_meja}. Using an instantaneous Langevin equation as a model, one typically studies the fluctuations of the injected power instead\cite{farago,bonaldi}, but its relation with the entropy production rate is subtle. Analysis of Langevin equations with explicit memory terms might offer insight and a new direction for approaching this problem.

As a concluding remark, we wish to briefly discuss the problematic issues that one is likely to encounter if FDT-type relations were sought for in regard to the memory functions\cite{vulpiani2}. This attempt may well be suggested by the form of the Langevin equations (\ref{circuit3}) and (\ref{circuit4}), essentially Langevin Equations with memory. Recall that FDT relations, in the presence of memory effects, assume causality conditions on the noise to be applicable, e.g. $\langle N(t) I(t-t') \rangle = 0$ for every $t' > 0$\cite{kubo,vulpiani}. However, from (\ref{newnoise}) and (\ref{newnoise_s}), one sees that $N(t)$, as well as $N_s(t)$, contain terms that are evidently causally related to $I(t-t')$ and $I_s(t-t')$, because they participated in the dynamical driving of the circuit at times earlier than $t-t'$ (one just needs take $kt_d$ to be sufficiently large). In other words, any random thermal fluctuation of the bath exerts its influence on the measuring apparatus both instantaneously as well as at later times, via the recycling of the feedback. A similar analysis can be performed also with the second type of setting. The approach that we have used here seems crystalline and controllable, being based on the sole assumption on the nature of the self correlations of the thermal fluctuations.

\ack{The research leading to these results has received funding from the European Research Council
under the European Community's Seventh Framework Programme (FP7/2007-2013) / ERC grant
agreement n¡ 202680. The EC is not liable for any use that can be made on the information
contained herein.}

\end{document}